\documentclass[9pt,twocolumn,twoside,lineno]{pnas-new}

\templatetype{pnasresearcharticle} 

\title{The structure of the blue whirl revealed}

\author[a,1]{Joseph D.\ Chung}
\author[a,1,2]{Xiao Zhang} 
\author[a]{Carolyn R.\ Kaplan}
\author[a,b]{Elaine S.\ Oran}

\affil[a]{Department of Aerospace Engineering, University of Maryland, College Park, MD 20742}
\affil[b]{Department of Aerospace Engineering, Texas A \& M University, College Station, TX 77843}

\leadauthor{Chung} 

\significancestatement{The growing worldwide demand to reduce emissions from combustion calls for development of alternative approaches for high-efficiency and low-emission combustion. Whereas fire whirls are known for their intense and disastrous threat to life and surrounding environments, their swirl properties and thus higher combustion efficiency imply an unexploited potential for highly efficient, low-emission combustion. Although fire whirls are inherently dangerous, they can be driven and so transition into an apparently benign state, the beautiful, small, laminar blue whirl. The blue whirl burns soot-free and consumes all fuel into product. Understanding the structure of the blue whirl serves as a fundamental first step to further controlling and utilizing it, suggesting new ways for fuel-spill remediation, reduced-pollution combustion, and fluid mechanics research.}

\authorcontributions{J.D.C., X.Z., C.R.K., and E.S.O. designed research, performed simulations, analyzed data, and wrote the paper.}
\authordeclaration{The authors declare no conflict of interest.}
\equalauthors{\textsuperscript{1} J.D.C. and X.Z. contributed equally to this work.}
\correspondingauthor{\textsuperscript{2}To whom correspondence should be addressed. E-mail: xiaoz@terpmail.umd.edu}

\keywords{blue whirl $|$ vortex breakdown $|$ triple flame $|$ combustion$|$ soot free } 

\begin{abstract}
The blue whirl is a small, stable, spinning blue flame that evolved spontaneously in laboratory experiments while studying, violent, turbulent fire whirls. The blue whirl cleanly burns heavy, liquid hydrocarbon fuels with no soot production, presenting a new potential way for low-emission combustion. It is reproducible, appears for a range of different fuels and initial conditions, is quiet, appears laminar, and has characteristics which led to the idea that it results from vortex breakdown in  whirling, reacting fluid. Since its discovery, considerable effort has been put into measurements, which have shown its temperature structure and sensitivity to the boundary layer near the surface. This has led to considerable speculation about the type of flames that comprise it. Simultaneously, there was a numerical effort to study its structure by performing simulations of vortex breakdown in gaseous reactive flows. The simulations described in this paper show that the stable blue whirl is composed of three different flame structures — a diffusion flame and a premixed rich and lean flame -- all of which meet in a fourth structure, a triple flame which appears as a whirling blue ring. In addition, the blue whirl structure emerges as the result of vortex breakdown in a swirling reactive flow, as evidenced in the simulation by a bubble mode that is usually invisible in the experiments but at the center of the whirl. This paper also presents the tool used for the study and discusses how this might be used for future investigations.
\end{abstract}

\dates{This manuscript was compiled on \today}
\doi{\url{www.pnas.org/cgi/doi/10.1073/pnas.XXXXXXXXXX}}

\begin{document}

\maketitle
\thispagestyle{firststyle}
\ifthenelse{\boolean{shortarticle}}{\ifthenelse{\boolean{singlecolumn}}{\abscontentformatted}{\abscontent}}{}



\begin{figure}[tbhp]
\centering
\includegraphics[width=.95\linewidth]{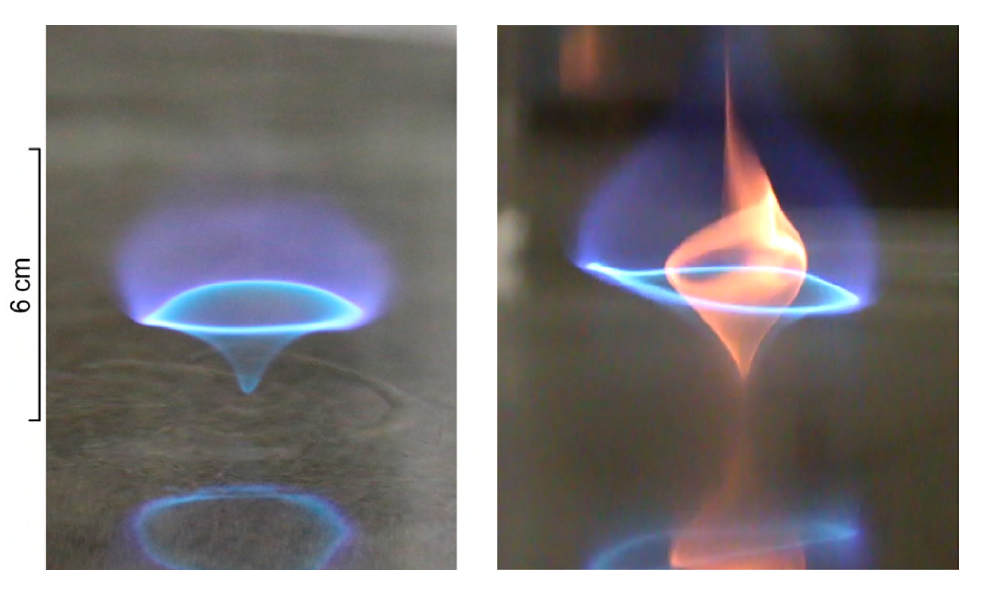}
\caption{a) A blue whirl. b) Slightly unstable blue whirl with yellow bubble in the middle, taken from \cite{xiao2016fire}.}
\label{fig:blue_whirl}
\end{figure}

\dropcap{A} blue whirl, shown in Fig.\ref{fig:blue_whirl}a,  is a small, soot-free blue flame that was  discovered serendipitously while performing experimental studies of fire whirls burning liquid hydrocarbon fuels on a water base \cite{xiao2016fire}.  Even though fire whirls are dangerous, violent, turbulent eddies of fire, they can be created in (relatively) safe, confined conditions for laboratory study (e.g., \cite{byram1962fire, emmons1967fire, lei2011experimental, hartl2016scaling}). Because fire whirls burn at higher temperatures \cite{grishin2005experimental} with higher burning rates \cite{morton1970physics, emmons1967fire, dobashi2016experimental} than their nonwhirling counterparts, preliminary studies \cite{xiao2016fire} were being performed to determine if it would be beneficial to use controlled fire whirls for practical purposes, such as oil-spill remediation. 

In the initial experiments \cite{xiao2016fire}, the blue whirl evolved spontaneously from a 1-m high fire whirl in a few seconds, as the whirling flame transitioned through a series of intermediate states. The result was that a noisy, turbulent, yellow fire whirl changed into a quiet, laminar, blue spinning flame. The intermediate states suggested the complex reactive-flow system was subject to the fluid dynamics instability, {\sl vortex breakdown}, which changes simple swirling flows into bubble, helical, or whirling  structures \cite{sarpkaya1971stationary,leibovich1978structure}. 

When a stable blue whirl becomes unstable for a short time, as shown in Fig.\ref{fig:blue_whirl}b, a yellow  bubble-like mode  appears in the center of the cone. This change in color, indicating the appearance of soot, led us to think of soot as a flow diagnostic that could be a hint of the otherwise invisible internal structure of the blue whirl. Since these initial experiments that demonstrated the blue whirl's existence, experiments have produced temperature maps \cite{thermalhariharan2019, bluehariharan2019}, $\text{OH}^*$ measurements \cite{bluehariharan2019}, scaling laws \cite{hu2019conditions}, ways to stabilize the blue whirl \cite{hu2019conditions}, and ways to create the blue whirl more easily {\cite{coenen2019observed}. A more complete list of experimental properties observed to date is given later in this paper. 

In addition, there has been considerable speculation about the fundamental flame structure of the blue whirl. There are two limits of laminar flames that are discussed quite separately in the literature. In a {\sl laminar premixed flame}, the flame front passes through premixed fuel and oxidizer, leaving behind the reaction products. The flame front is driven by expansion due to heat release from the reactions and physical diffusion processes, such as thermal conduction, molecular diffusion, and radiation transport. There can be fuel-lean, stoichiometric, and fuel-rich premixed flames. This is to be contrasted to a {\sl laminar diffusion flame}, in which the fuel and oxidizer are initially separated and mix by physical diffusion processes. In this case, the rates of reactions are controlled by diffusion and the flame is said to be ``diffusion limited.''  

Thus a fundamental  question for combustion theory that was posed by the blue whirl is: {\sl What is its flame structure of the blue whirl?} Is it a premixed flame or a diffusion flame, or some combination? There was considerable speculation, including that made in \cite{xiao2016fire}, but no definitive answers.

Parallel to experiments, there have been computational and theoretical efforts to simulate fire whirls and the evolution to a blue whirl. Simulating a realistic fire whirl is expensive computationally because of the very wide range of space and time scales involved. Simulating a blue whirl would mean either simulating a fire whirl subject to vortex breakdown, or finding a way to go more directly to blue-whirl conditions. At the beginning of the simulation effort, we did not know which approach, or whether a combined approach, would work best.

This paper presents the first results of unsteady, three-dimensional (3D) numerical simulations that examine vortex breakdown in a reactive flow that leads to a blue whirl. It reveals the flame and flow structure of the blue whirl through a series of numerical diagnostics,  relates the results to prior experiments, and suggests a path forward for both future experiments and simulations to examine and potentially use this new, soot-free flame structure.

\section*{Approach to the problem}

The approach to creating the simulations followed this development path:

\begin{enumerate}[wide, labelwidth=!, labelindent=0pt]

\item Simulate vortex breakdown in a nonreactive gas in order to observe the modes induced by vortex breakdown as they evolve in a gaseous reactive flow. This led to the development of the low-Mach-number algorithm \cite{zhang2018barely} described in more detail at the end of the paper.

\item Develop a chemical-diffusion model (CDM) that reproduces features of a diffusion flame as well as a premixed flame, and find parameters for it suitable for n-heptane \cite{chung2019low} used as the fuel in the original experiments.

\item Simulate a fire whirl and ensure that the resulting flow and properties are consistent with observations. This required generalizing the low-Mach-number algorithm so that it is able to simulate reactive flow, with energy release and species conversion \cite{zhang_chung_kaplan_oran_2019}. 

\item  Simulate {\sl reactive vortex breakdown} \cite{chung_zhang_kaplan_oran_2019}, as it would occur when the swirling gas consists of an ignited mixture of fuel and air. The conditions should be similar to those that produced the experimentally observed blue whirl.

\item Use the new numerical model and the general initial conditions of the experiment to reproduce the blue whirl numerically. 

\end{enumerate}

\noindent
Some of the important experimentally observed properties of the blue whirl are: 

\begin{itemize}[leftmargin=*, topsep=8pt,itemsep=4pt,partopsep=4pt, parsep=4pt]

\item  Initial and boundary conditions (e.g., size, fuel flow) leading to a stable blue whirl in n-heptane \cite{xiao2016fire}.

\item  Swirl levels (vane inclination angles) leading to a stable blue whirl in ethanol and heptane \cite{coenen2019observed}.

\item  Temperature measurements \cite{thermalhariharan2019, bluehariharan2019}.

\item  OH* measurements \cite{bluehariharan2019} 

\item  Regimes of circulation $\Gamma$ vs heat release $\dot q$ in which the blue whirl exists \cite{hu2019conditions}.

\end{itemize}
%
%
%
%
%

\noindent 

Experimental measurements have given us considerable information about the formation conditions and the final state of the blue whirl. Nonetheless, the flow and flame structure are still not certainly defined: the blue whirl moves around and is difficult to diagnose. This led us to believe that a full numerical simulation, from a fire whirl to a blue whirl, would be needed to tell us what the blue whirl really is. Such a simulation capability could also be used with experiments to study fundamental questions, such as whether the blue whirl scales or how to create it more directly without going through the full, dangerous fire whirl state. This is the motivation that leads to a computational ``hunt’’ for the blue whirl, in which we first developed the method and then used the simulations to explore the effects of varying boundary conditions and inflow fuel velocities.

The 3D unsteady numerical simulation described in this paper is one of many simulations carried out in which boundary and and fuel inflow conditions were successively varied until a flame structure appeared that was in qualitative and even quantitative agreement with the observed blue whirl. Such a simulation provides the full picture of the dynamics of the primary variables (density, momentum, and energy, and species conversion from fuel to product in background air), given a set of  initial and boundary conditions and material properties. The exact conditions for the particular simulation  analyzed below are discussed later in the paper. In addition to  primary variables, we use derived quantities (e.g., temperature and pressure), and flow and reaction diagnostics (e.g., flow streamlines, flame index, heat release rate, etc) to analyze the flow structure. The primary difference in physical conditions from the reported experiments is that the process of heptane evaporation was bypassed by assuming a small forced inflow of pure heptane gas (here 371 K) at the bottom of the domain. Recent experiments have shown the blue whirl can be obtained from gaseous fuel injection, which verifies our approach of not including fuel evaporation.


\section*{The Flame Structure Revealed}

Figure \ref{fig:schematic}a is a volume rendering of the heat release rate from the final result of the blue-whirl simulation effort. Figure \ref{fig:schematic}b is a schematic diagram that summarizes the result. It  is posed next to Fig.\ \ref{fig:schematic}c, the observed blue whirl. We see now that the blue whirl is composed of four types of flames. The lower part of the blue whirl is a rich premixed flame, and the purple crown is a diffusion flame. What cannot be seen easily in the laboratory experiments is the lean premixed flame surrounding the purple haze, that is, the upper region just outside of the diffusion flame. The bright blue ring is where the three types of flames meet, which is a triple flame. The interpretations presented in the schematic diagram in Fig.\ \ref{fig:schematic} are derived from data extracted from simulations in which an initial flow structure was given and allowed to evolve to a point where the basic blue-whirl structure no longer changed significantly in time.

\begin{figure*}[tbhp]
\centering
\includegraphics[width=11.4 cm]{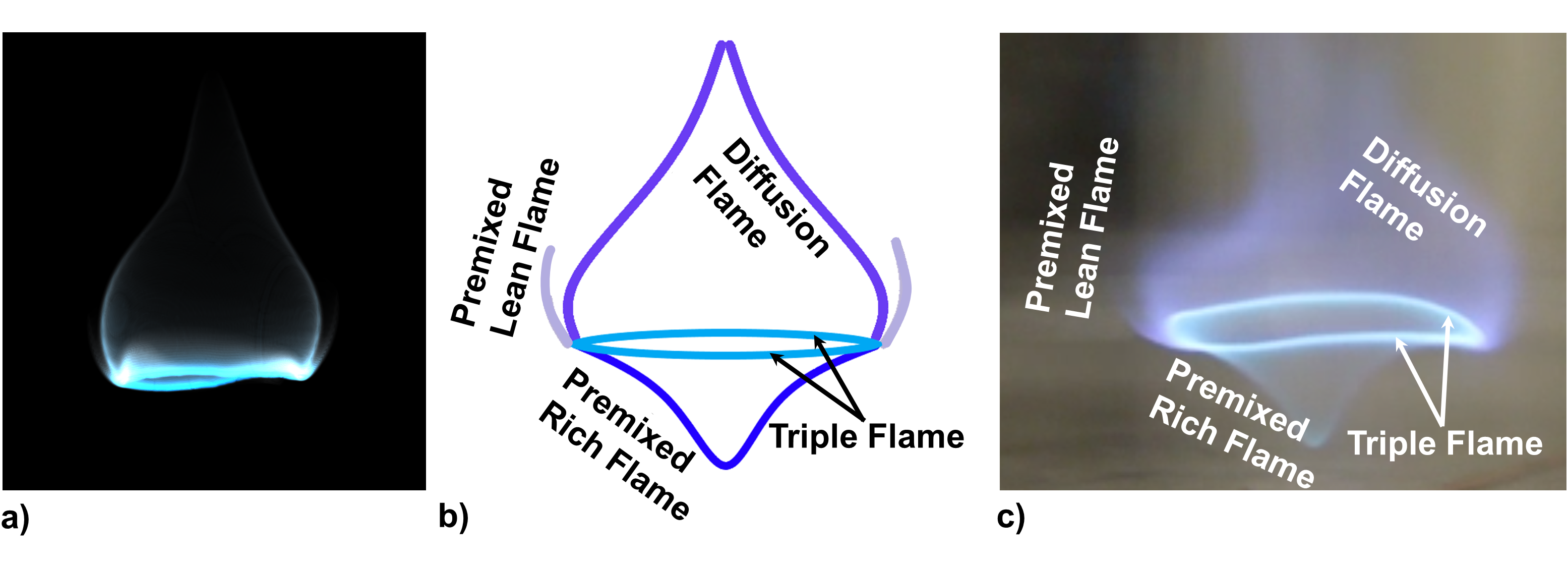}
\caption{The flame structure of the blue whirl. a) Volume rendering of the heat release rate from the numerical simulation described in the text. b) Schematic diagram that summarizes a final result of the blue whirl simulation. c) Observed blue whirl.}
\label{fig:schematic}
\end{figure*}

The computational setup, including the mesh and the initial and boundary conditions, are shown in Fig.\ \ref{fig:mesh}. The domain is a cube with sides that are $30\,\rm cm$ long. The upper boundary is an outflow condition and all other boundaries are non-slip, adiabatic walls. Heptane vapor is injected within a $0.9\,\rm cm$ diameter at the center of the bottom wall with a constant velocity of $5.8\,\rm cm/s$ and at the evaporation temperature of heptane at $1\,\rm atm$, $371\,\rm K$. Circulation is applied by forcing air through the four corners with a speed of $40\,\rm cm/s$ along slits which are $5\,\rm cm$ wide. Radial inflow is introduced by forcing air with a velocity of $60\,\rm cm/s$ through a $1.4\,\rm cm$ high and $16\,\rm$ wide region along the lower portion of the walls. The interior domain is initialized with quiescent air at $1\,\rm atm$ and $298\,\rm K$ with a column of hot product gas that is $1\,\rm cm$ in diameter and $10\,\rm cm$ high just above the fuel inflow for ignition. 

As shown in Fig.\ \ref{fig:mesh}b, the simulations were performed on a 3D mesh which concentrated a fine grid along the center to cover the region of a blue whirl. The fine grid region is $10\,\rm cm$ in width, $10\,\rm cm$ in depth, and $10\,\rm cm$ in height.  For the simulation shown, the width of the finest cell size in the center region was $0.01465\,\rm cm$, corresponding to 5 levels of refinement from the coarsest cells at the edge of the domain. The numerical method and computational resources required for solving the reactive Navier-Stokes equations that produces the results shown is described in more detail at the end of the paper. 

\begin{figure}[tbhp]
\centering
\includegraphics[width=.9\linewidth]{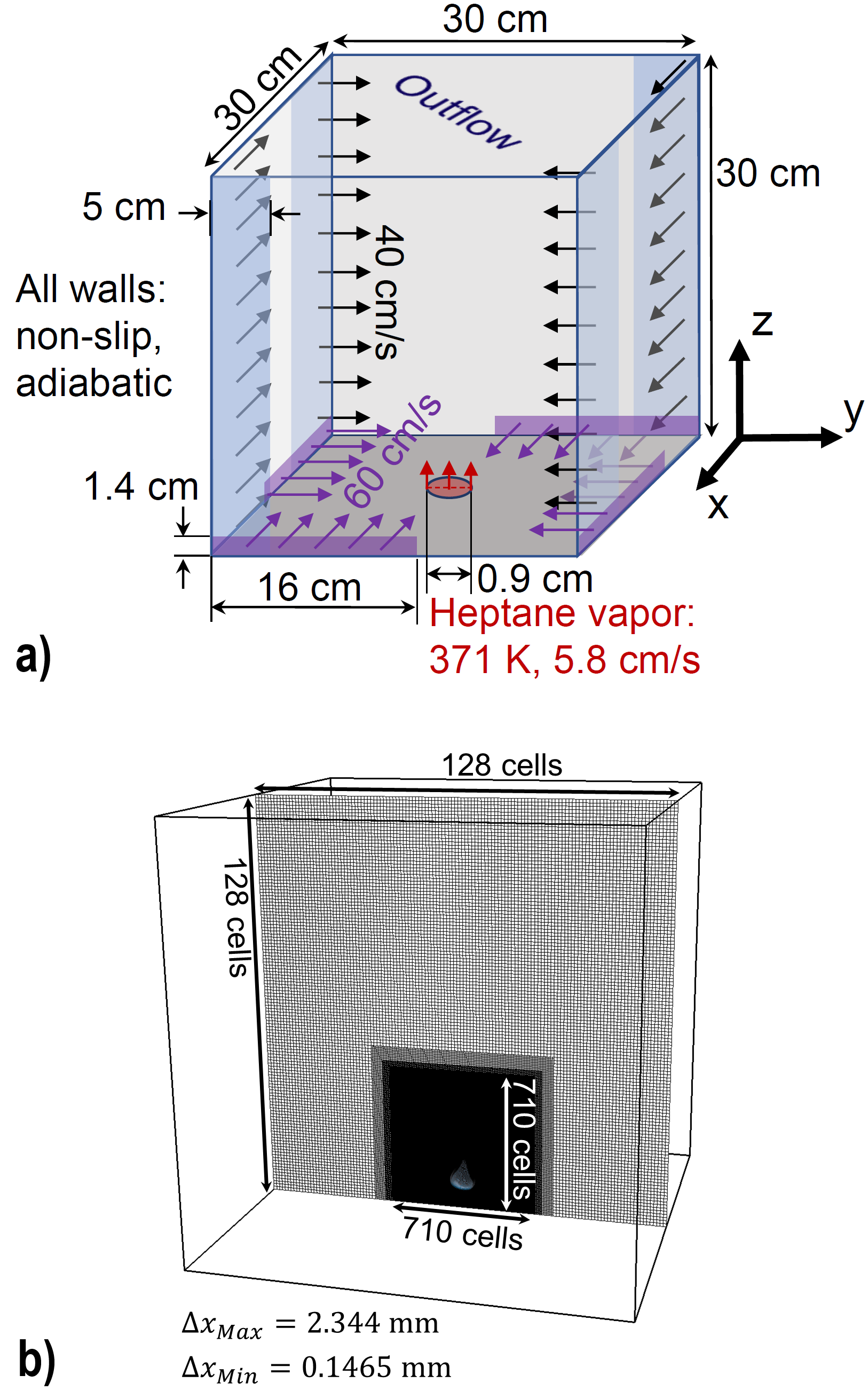}
\caption{Computational setup. a) Schematic of the computational domain and the boundary conditions. b) A center slice of the 3D computational mesh. The mesh is composed of cubical control volumes. The width of the control volume in each level of refinement is half the width of the coarser level. The mesh is refined around the blue whirl which is shown as a volume rendering of the heat release rate.}
\label{fig:mesh}
\end{figure}

Figure \ref{fig:OH} is a comparison between the blue whirl experiment \cite{bluehariharan2019}  and the simulation. The luminosity in Fig.\ \ref{fig:OH}a shows the experimental OH* concentration \cite{bluehariharan2019}  which indicates the intensity of the reaction. For the simulation result, this is indicated by the 3D volume rendering of the heat release rate shown in Fig.\ \ref{fig:OH}b (now readjusted in greyscale in contrast with Fig.\ \ref{fig:schematic}a). Bright regions indicate stronger reaction and darker regions indicate weaker reaction. Both the experiment and simulation show that a significant amount of combustion occurs within the blue ring. The simulation result shown here agrees well with the experimental measurement in terms of curvature of the reaction regions and distribution of the reaction.  

\begin{figure}[tbhp]
\centering
\includegraphics[width=.9\linewidth]{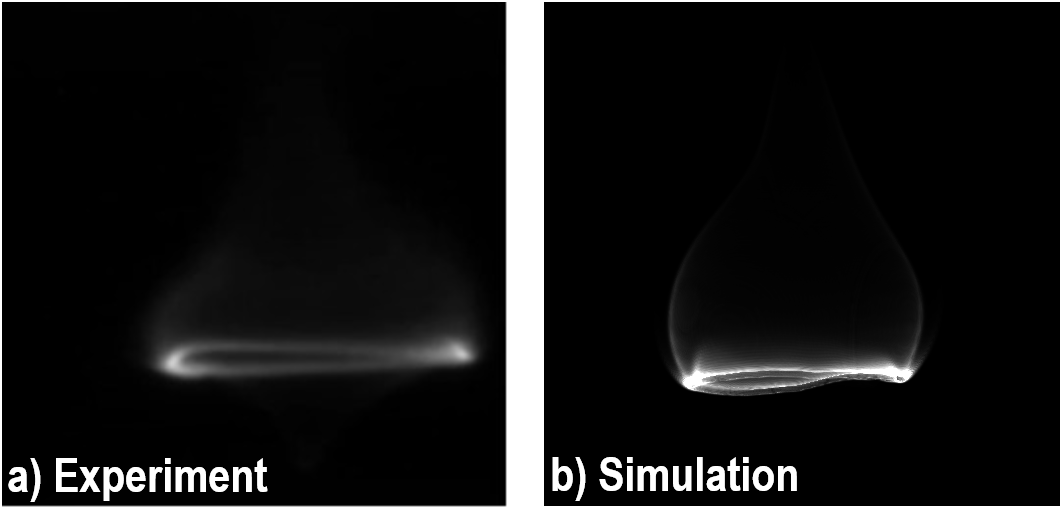}
\caption{Comparison of a) experimental OH* concentration measurement  (taken from Fig. 8a  in \cite{bluehariharan2019}) with b) 3D volume rendering of heat release rate in the simulation. The volume rendering is taken from the side view. }
\label{fig:OH}
\end{figure}

Figure \ref{fig:flame_index}a is a map of the flame index \cite{yamashita1996numerical}, $ I_f ~=~ {\displaystyle  {\nabla Y_{\rm Fuel} \cdot \nabla Y_{\rm Ox}} /{ \displaystyle {| \nabla Y_{\rm Fuel} | } {| \nabla Y_{\rm Ox} | }}  } $, where $Y_{\rm Fuel} $ and $Y_{\rm Ox} $ are computed values of the mass fraction of fuel and oxidizer, respectively. $I_f > 0$ is a premixed flame  and   $I_f < 0$  is a diffusion flame. Fig.\ \ref{fig:flame_index}b is the corresponding map of equivalence ratio, $\phi$, and Fig.\ \ref{fig:flame_index}c shows temperature. Contours of heat release rate are superimposed on each figure. 

\begin{figure*}[bh]
\centering
\includegraphics[width=11.4 cm]{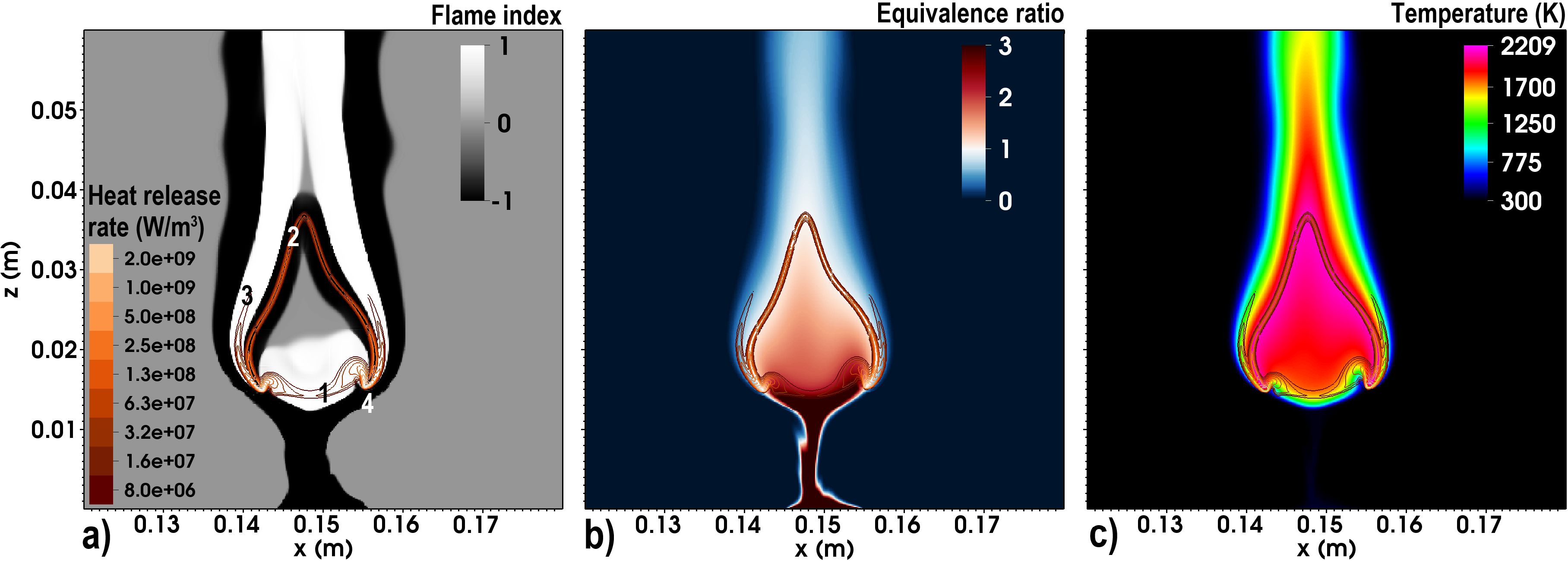}
\caption{Slices through the center of the computational domain and parameters selected for combustion diagnostics. a) Flame index. b) Equivalence ratio. c) Temperature. Contours of heat release rate are superimposed on top to indicate reaction regions. Slices are shown for a zoomed in region which is $8\rm\ cm$ wide.}
\label{fig:flame_index}
\end{figure*}

In Fig.\ \ref{fig:flame_index}a, region 1 has positive $I_f$, and in Fig.\ \ref{fig:flame_index}b, region 1 has equivalence ratio larger than 1. Taken together, the heat-release rate in region 1 corresponds to a premixed fuel-rich flame. By similar reasoning, the heat release rate in region 2 corresponds to a diffusion flame, because region 2 has negative $I_f$ and an equivalence ratio of 1. The heat release rate in region 3 corresponds to a premixed fuel-lean premixed flame because region 3 has positive $I_f$ and equivalence ratio less than 1. Region 4 is where the three flames meet and is the triple flame (or blue ring in the experiments). It has the most intense heat release, which is consistent with the $\text{OH}^*$ experimental measurements \cite{bluehariharan2019} as mentioned earlier. The temperature map, Fig.\ \ref{fig:flame_index}c, shows that the hottest regions are the diffusion flame in the purple crown, in agreement with the experimental measurements \cite{thermalhariharan2019, bluehariharan2019}, and the region at the bright blue ring (that could not be measured in the experiment). Fig.\ \ref{fig:flame_index} shows a gap between the flame and bottom surface, again consistent with the experimental observations \cite{xiao2016fire}. Analysis of the composition of the data at the top of the computational grid shows that essentially all of the fuel is consumed in the blue whirl and only hot product and air exit the computational domain. 

The flow structure shown confirms and elaborates on experimental observations. Figure \ref{fig:streamlines} is a composite showing (a) axial and (b) tangential velocities and (c) streamlines, again all with superimposed heat release rate, as well as (d) a profile of the tangential velocity through a slice at the bottom of the blue whirl. The evolution to this structure from the initial conditions (not shown here) in the simulation show the development of a fire whirl which undergoes vortex breakdown, leading to the typical bubble-mode seen in Fig.\ \ref{fig:streamlines}. Below the flame, the axial velocity map shows  a jet-like velocity profile and the tangential velocity map shows high peak velocity and a narrow vortex core. Figure \ref{fig:streamlines}d, a profile along the white dashed line in Fig.\ \ref{fig:streamlines}b, shows that there is a sharp gradient in the core. The axial velocity Fig.\ \ref{fig:streamlines}c shows a negative velocity region at the lower part of the flame. Taken together with the streamlines in Fig.\ \ref{fig:streamlines}a, we again see the structure of a vortex-breakdown bubble inside the flame.

\begin{figure*}[!ht]
\centering
\includegraphics[width=17.8cm]{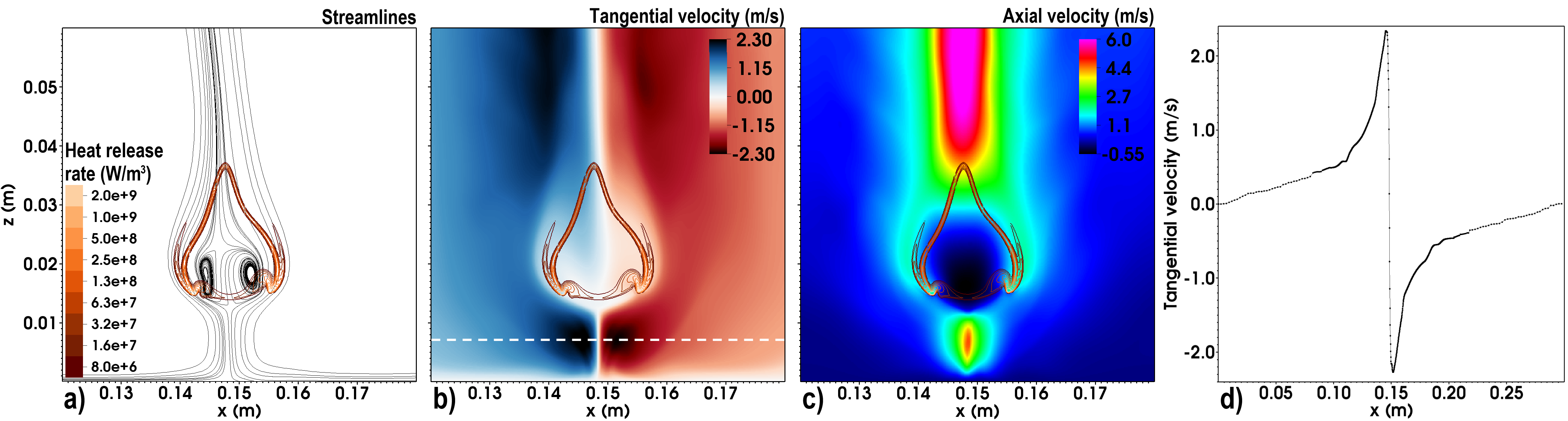}
\caption{Slices through the center of the computational domain and values selected for flow diagnostics. a) Streamlines. b) Tangential velocity. c) Axial velocity. Contours of heat release rate are superimposed on top to indicate reaction regions. Slices are shown for a zoomed in region which is $8\,\rm cm$ wide. d) Line plot of tangential velocity taken below the blue whirl from the white dashed line in b), shown for the entire width of the computational domain.}
\label{fig:streamlines}
\end{figure*}

The streamlines show that the vortex rim is inside what we see as the blue ring. This is in qualitative agreement with the experimental results shown in Fig.\ \ref{fig:blue_whirl}b, where the recirculation zone illuminated by the soot pattern is inside the blue rim. In the upper portion of the flame, above the bubble, the axial velocity map shows that the flow is accelerating. The upper portion of the  tangential velocity map shows the flow recovering the vortex structure as it leaves the bubble.

Finally, we use information from the flow streamlines in Fig.\ \ref{fig:3D_streamlines} superimposed on a 3D map of heat release rate (yellow structure in the figure) to show how air from the boundary layer is introduced into the flame. The streamlines are colored by the local temperature of the flow. The four streamlines start at 5 cm from the center of fuel injection on an $x-y$  plane. The streamlines in Fig.\ \ref{fig:3D_streamlines}a and Fig.\ \ref{fig:3D_streamlines}b originate at two different heights from the lower boundary, 0.5 mm and 2.0 mm, respectively. 

\begin{figure}[tbhp]
\centering
\includegraphics[width=.8\linewidth]{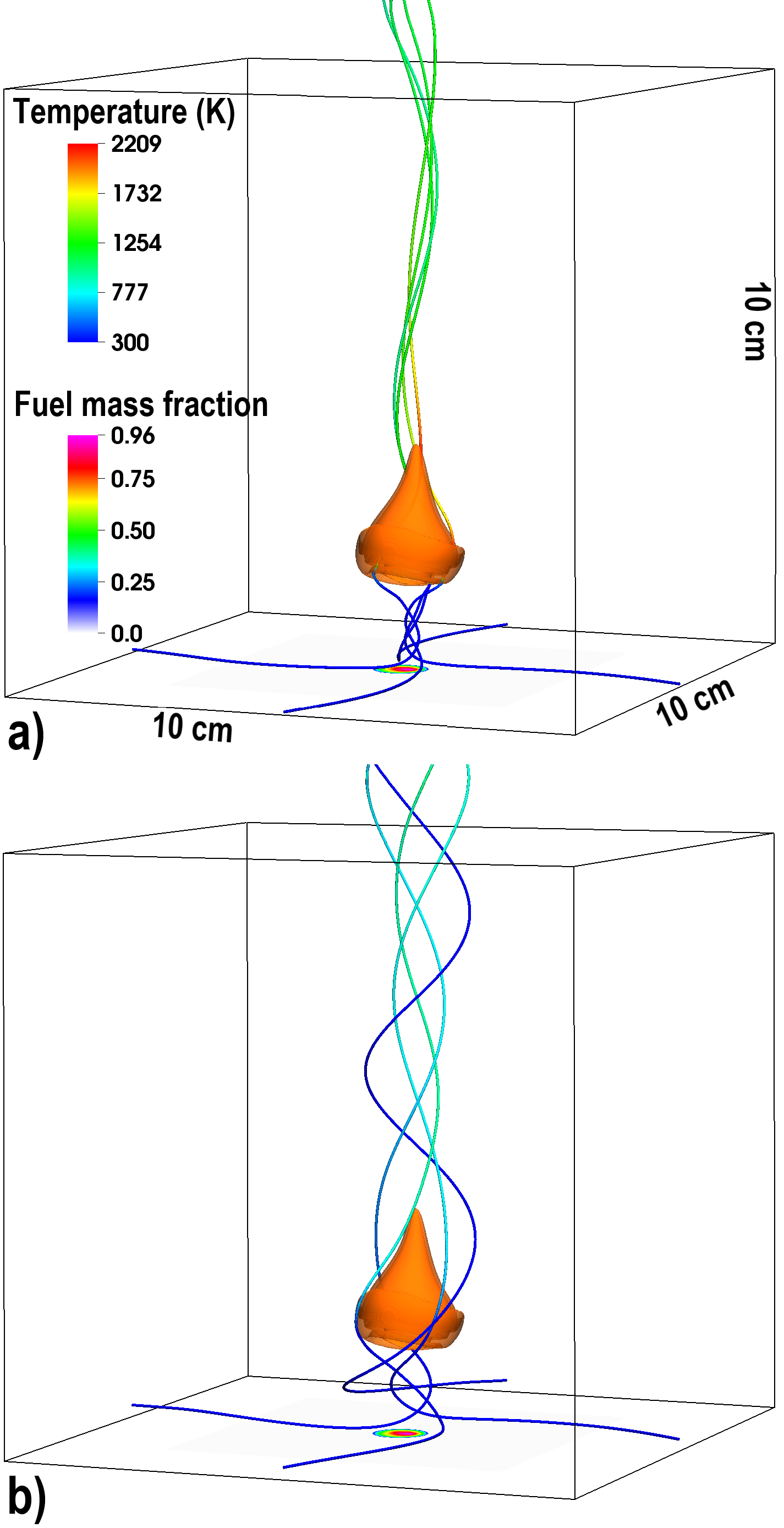}
\caption{Streamlines superimposed on a 3D heat release rate isocontour of $3\rm\ MW/m^3$.  a) Streamlines which originate at $0.5\rm\ mm$ from the lower boundary. b) Streamlines which originate at $2.0\rm\ mm$ from the lower boundary. The streamlines are colored by the local temperature of the flow. A 2D map of fuel mass fraction along the bottom boundary is shown, indicating the region of fuel inflow. The box indicates the region of mesh refinement.}
\label{fig:3D_streamlines}
\end{figure} 

First, from Fig.\ \ref{fig:3D_streamlines}b, we see that air from the higher portion of the boundary layer maintains a low temperature of 300 K even after moving around the flame. This shows that  air from the upper part of the boundary layer is not involved in the combustion process.  (This is also consistent with the experiments in which you can put your hand right up to the flame on the sides and it does not feel hot.)

The story is different, however, with air from the lower part of the boundary layer, shown in Fig.\ \ref{fig:3D_streamlines}a, depending  on the height at which the flow reaches the upward draft and is pulled into the flame. Air from very close to the bottom boundary, as shown here, first encounters the heptane vapor and is mixed due to the strong circulation below the bubble. This forms the rich premixed flame conditions seen at the bottom of the bubble. Then most of the residual, unburned fuel and product is pulled into the bubble, which is shown in bubble mode simulations. This region inside the bubble creates the fuel rich region which feeds the diffusion flame in the crown. 

Meanwhile and simultaneously, air from higher in the boundary layer, here between the bottom (e.g., the 0.5 mm height) and the 2.0 mm height, is drawn upwards and flows around the entire structure. Outside air and the residual fuel from inside the bubble set up a diffusion flame bordering the crown. A small amount of fuel also leaks outside of the bubble and burns with the outside air to form a very lean premixed flame outside of the crown. 

All of these flames -- the rich premixed flame, the diffusion flame, and the lean premixed flame -- must come together at some point, and this occurs at the blue ring surrounding the bubble. This blue ring has the most intense heat release and it burns as a triple flame.

\section*{Summary and Discussion Points}

In summary, a 3D time-dependent solution of the Navier-Stokes equations coupled to  a model for chemical energy release and species conversion from fuel to product for heptane gas was used to simulate conditions that lead to the blue whirl and reveal its flow and flame structure (Fig.\ \ref{fig:schematic}). There are several elements of the physical results that should be discussed before the numerical  model is described in more detail. 

The result was achieved by using the experimental conditions as a starting point and then varying the physical conditions represented in the calculations until the blue whirl appeared. Notable differences between the experiments reported and the simulations include: the shape of the external container;  fuel injection vs fuel evaporation; self-determining inflow boundary conditions vs forced air inflow.   

Experiments performed in square containers have been reported to produce blue whirls. Thus we know the blue whirl does not depend on the external shape of the container. We know, however, that the blue whirl is sensitive to the inflow boundary conditions, and that it is more easily formed when the inflow conditions are as smooth and laminar as possible. Finally, this structure and its flow properties provide an excellent starting point for examining some of the fundamental questions related to the blue whirl, such as how and whether it might scale to larger sizes and whether it can be made directly without going through the fire whirl state. 

\section*{Implications of this Work}

The blue whirl is {\sl at least} a curious phenomenon that has many intriguing aspects. The most curious aspect is that it evolves spontaneously and presents itself as a stable state  persisting until all of the fuel is burned. The second curiosity was that it is laminar and burning soot free, whereas the initial state was sooty, turbulent, and noisy.  A third curiosity was that in the experiments, it was not burning a gas, but a liquid hydrocarbon sitting on a water surface. Further experimentation revealed more features, such as its averaged temperature profile and its sensitivity to the boundary layer. Added to all of this was that it was very beautiful, both in its stable state, as a spinning blue top-like flame, and when it went slightly unstable, perhaps revealing some of its inner structure. The route to its formation and its transient unstable states implied its relation to the fluid phenomenon of vortex breakdown and the various states that evolve from this instability.

A recurring question, however, was whether the blue whirl could be useful in any way for efficient combustion with no soot formation. This involves questions such as: Can it be formed under controlled conditions more directly and without going through the fire whirl state? Can the size be controlled? Can it be made larger or smaller? Is there a scaling that can be used? Other, perhaps more far out questions, were: Can it be made without the confining walls? Can multiple blue whirls be made and work together? Could it be part of a combustor or a propulsion device? The lure of being able to burn any liquid hydrocarbon efficiently and cleanly is extremely attractive.

None of these questions can be answered easily until we at least understand the structure and dynamics of the flame and have a tool through which we can easily explore some of these questions. This paper describes a first step: a tool that can be used to explore and test the phenomenon, and how it has been used to reveal the blue whirl structure. 

\section*{Methods: Computational Technology}

To be able to compute the blue whirl, new algorithms and a new Navier-Stokes solver for low-Mach-number flows were developed, refined, and implemented on a variably spaced 3D adaptive grid. For these computations however, static but not dynamic adaption was used. The underlying concept of the low-Mach-number algorithm is based on BIC-FCT, the barely implicit correction to flux-corrected transport, and is referred to here by the same name \cite{patnaik1987barely}. The underlying fluid solver is based on flux-corrected transport \cite{boris1993lcpfct,zalesak1979fully,devore1998improved}. Then the solution is  modified to filter out high-frequencies of the sound-wave spectrum, thus removing the computational restriction imposed by sound waves, but not removing all compressibility effects. To do this, a pressure correction term, $\delta P$, is computed and used to modify the energy and momentum  equations and enable long computational time steps. 

One result of this long time step, however, is that the effects of nonsimultaneity in the various parts of the solution are exacerbated. To deal with these, a conservative monotone filter was applied repeatedly to damp unphysical fluctuations as they arose. The exact procedure and results of a number of one-dimensional (1D), two-dimensional (2D), and 3D accuracy and resolution tests were applied to a series of test flows \cite{zhang2018barely}. The result is an extremely stable and robust code capable of solving problems with many types of boundary conditions \cite{zhang2018barely, zhang2019comparison}. Extensions to reactive flows, with physical diffusion effects, chemical energy release, and species conversion have been reported in \cite{chung2019low, zhang_chung_kaplan_oran_2019}.

The effects of diffusion and chemical reactions with heat release for heptane and air mixtures were incorporated into the solver using the chemical-diffusive model (CDM)  \cite{kessler2010simulations, kaplan2019chemical}.  The CDM is a way to represent chemical reactions and heat release while maintaining constraints, such as: flames and detonations propagate at the correct speed with the correct length scale and the temperatures of the product gases are correct. The model is derived by proposing a mathematical form with constants that are fit to maintain the constraints. The constants are found from an iterative calibration procedure that incorporates fundamental principles of combustion and diffusion processes. The values of the constraints can come from experiments or detailed chemical mechanisms. In the case here for heptane, we calibrated the CDM \cite{kaplan2019chemical, chung2019low} using the flame and diffusion properties from a more detailed chemical mechanism representing heptane-air combustion \cite{lu2006linear}.

The adequacy of the numerical resolution was tested by increasing the levels of refinement in the blue-whirl region until there were no changes to the flow and flame structure. Refinement required for a premixed flame is reported in \cite{kessler2010simulations}. In this computation, there are enough computational cells  within the flame thickness to give at most an 8\% difference between the constraint data and the CDM prediction. In the rich and lean flame regions, this difference is smaller because there are more cells within the flame thickness since nonstoichiometric flames are thicker. Resolution required for diffusion flames is a cell size of $0.07\,\rm cm$ or smaller. This was determined by solving 2D counter-flow diffusion flames. 

The computational search for the blue whirl took its lead from the experiments. Many computations with variations in geometrical, physical, and computational parameters were required to find this solution shown above. Critical elements in finding the solutions consisted of determining the appropriate air and fuel inflow geometry and the inflow rate of air and fuel to allow vortex breakdown to occur, the flame to lift away from the bottom surface, and the blue whirl to form. We started with simulations of fire whirls. Then, we chose air inflow conditions at the corners and fuel mass flow rate to be similar to those measured for the blue whirl and obtained vortex breakdown. The corner air and fuel inflow geometry and flow rates were iteratively varied until we obtained a flame which was lifted from the bottom surface while maintaining vortex breakdown. The radial inflow geometry and flow rate were iteratively varied until the lifted flame was stabilized. Then, the mesh was refined to four levels, corresponding to a cell size of $0.0293\,\rm cm$, resulting in the blue whirl structure. The mesh was further refined to five levels which did not drastically change the flow or flame structure, demonstrating that the solution is well resolved. With three levels of refinement, this computation covered $15\,\rm s$ of physical time and with four and five levels of refinement, the computation covered $0.6\,\rm s$ of physical time. The final mesh with five levels of refinement contains 410 million cells. The computation overall took $600,000$ CPU hours on 40 Dell Poweredge C8220 nodes using dual Intel Ivy Bridge E5-2680v2 processors running at 2.80 GHz with 20 cores per node.

\acknow{The authors would like to thank the National Science Foundation Grant CBET 1839510, the Army Research Office Grant W911NF1710524, Minta Martin Endowment Funds in the Department of Aerospace Engineering at the University of Maryland, the Glenn L. Martin Institute Chaired Professorship and the A.\ James Clark Distinguished Professorship at the A.\ James Clark School of Engineering at the University of Maryland, the TEES Distinguished Professorship of Texas A\& M University for their support of this work. The authors are extremely grateful to Michael Gollner and Sriram Bharath at the University of Maryland for their advice and help and many others who will be included in the final version.  The computations in this study were performed on the University of Maryland, Deepthought2 cluster and Thunder from the Air Force Research Laboratory.}

\showacknow{} 

\bibliography{BW}

\end{document}